# Quantum Fast Poisson Solver: the algorithm and modular circuit design

Shengbin Wang[a], Zhimin Wang[a*], Wendong Li[a], Lixin Fan[a], Zhiqiang Wei[b] and Yongjian Gu[a†]

[a] *Department of Physics, College of Information Science and Engineering, Ocean University of China, Qingdao 266100, China*
[b] *Department of Computer Science and Technology, College of Information Science and Engineering, Ocean University of China, Qingdao 266100, China*

ABSTRACT

The Poisson equation has applications across many areas of physics and engineering, such as the dynamic process simulation of ocean current. Here we present a quantum Fast Poisson Solver, including the algorithm and the complete and modular circuit design. The algorithm takes the HHL algorithm as the template. The controlled rotation is performed based on the arc cotangent function which is evaluated by the Plouffe's binary expansion method. And the same method is used to compute the cosine function for the eigenvalue approximation in phase estimation. Quantum algorithms for solving square root and reciprocal functions are developed based on the non-restoring digit-recurrence method. These advances make the algorithm's complexity lower and the circuit-design more modular. The number of the qubits and operations used by the circuit are $O(d\log^2(\varepsilon^{-1}))$ and $O(d\log^3(\varepsilon^{-1}))$, respectively. We demonstrate our circuits on a quantum virtual computing system installed on the Sunway TaihuLight supercomputer. This is an important step toward practical applications of quantum Fast Poisson Solver in the near-term hybrid classical/quantum devices.

## I. INTRODUCTION

Quantum computing exploits quantum mechanical features, especially the quantum superposition and entanglement, to solve efficiently the problems intractable for classical computing [1,2]. Applications of quantum computing are based on quantum algorithms that mainly include Shor's factoring algorithm [3], Grover's searching algorithm [4] and the HHL algorithm [5], *etc*. HHL algorithm has been being the center in the research of quantum machine learning [6] and is also inspirational in many ways for finding quantum algorithms solving differential equations [7-11].

The Poisson equation is a widely used linear differential equation, which plays a key role across many areas of physics and engineering. For instance, when you simulate the dynamic process of ocean current, the Navier-Stokes equations are a good start to calculate the velocity field of the current, but notoriously hard to solve; while using the well-studied vortex-in-cell method, the velocity field can be evaluated rather easily from a vector potential which satisfies the Poisson equation [12,13]. So solving the Poisson equation constitutes the most computationally intensive part of the current simulation. In the classical algorithms, several kinds of Fast Poisson Solver has been developed to deal with it [14-16].

Recently Cao et al. proposed a quantum algorithm for solving the *d*-dimensional

[*] Email address: wangzhimin@ouc.edu.cn
[†] Email address: guyj@ouc.edu.cn

Poisson equation based on the HHL algorithm [11,17]16], which achieves an exponential speedup against any classical algorithms in terms of dimension of the Poisson equation. They presented in principle a scalable quantum circuit for the algorithm, of which the number of qubits and quantum operations used are proportional to, respectively, $\max\{d, \log_2 \varepsilon^{-1}\} \cdot (\log_2 \varepsilon^{-1})^2$ and $\max\{d, \log_2 \varepsilon^{-1}\} \cdot (\log_2 \varepsilon^{-1})^3$ within the error of $\varepsilon$. However, there still remains a certain amount of work to do to establish a quantum Fast Poisson Solver that is implementable on a near term middle-scale quantum device. Specifically, the most complex parts of Cao's algorithm are the rotation-angle calculation in the controlled rotation and the eigenvalue approximation in the phase estimation, but circuits design for both parts were lack in their work. In fact, the dominant cost results from computing the reciprocal, trigonometric and inverse trigonometric functions through the Newton iteration and Taylor expansion methods. That need a number of iteration steps, and in each iteration step several multiplication and division operations are calculated reversibly. So large number of auxiliary qubits and quantum operations are required.

In the present work, we try to redesign the algorithm for solving the Poisson equation, and present a complete and modular quantum circuit to establish a quantum Fast Poisson Solver. More specifically, first we find a new way of implementing the controlled rotation based on the arc cotangent function which is evaluated by the Plouffe's binary expansion method [18]. Secondly, we revise the Plouffe's binary expansion method to be able to compute the cosine function in a simple recursive way, and use it to approximate eigenvalues for the phase estimation. Thirdly, we develop quantum algorithms for solving the reciprocal and square root functions based on the non-restoring method. This method is a kind of digit recurrence method in classical algorithms [19,20]. All these developments would reduce the present algorithm's complexity and make the circuit more modular and implementable. Finally, we demonstrate our circuits on a quantum virtual computing system installed on the Sunway TaihuLight supercomputer.

This paper is organized as follows. In Sec. II the overview of the problem is described. In Sec. III we describe the quantum circuit in detail, including the implementations of each module, algorithm complexity and error analysis. Sec. IV shows the demonstration results of the present circuits on the quantum virtual system. Finally, conclusions and outlook of the present work are discussed in Sec. V.

## II. OVERVIEW OF THE PROBLEM

The problem is solving the Poisson equation over a unit rectangular or cube domain with Dirichlet boundary conditions. Let us first focus on the one-dimensional Poisson equation which could be described by the following equations,

$$-\frac{d^2 v(x)}{dx^2} = b(x), x \in (0,1),$$
$$v(0) = v(1) = 0,$$
(1)

where $b(x)$ is a given smooth function representing the charge or velocity distribution in different questions, and $v(x)$ the solution to compute. Using the central difference approximation to discretize the second derivative, Eq. (1) could be written in finite difference form as

$$h^{-2}(-v_{i-1} + 2v_i - v_{i+1}) = b_i, i = 1, 2..., N-1,$$
$$v_0 = v_N = 0,$$
(2)

where the number of discrete points is $N+1$ and the mesh size $h$ equals to $1/N$. Then we have $N-1$ linear equations which can be expressed as follows,

$$A \cdot \begin{pmatrix} v_1 \\ v_2 \\ \vdots \\ v_{N-1} \end{pmatrix} = h^{-2} \begin{pmatrix} 2 & -1 & & 0 \\ -1 & \ddots & \ddots & \\ & \ddots & \ddots & -1 \\ 0 & & -1 & 2 \end{pmatrix} \cdot \begin{pmatrix} v_1 \\ v_2 \\ \vdots \\ v_{N-1} \end{pmatrix} = \begin{pmatrix} b_1 \\ b_2 \\ \vdots \\ b_{N-1} \end{pmatrix}. \quad (3)$$

Now the problem of solving Poisson equation transfers into solving the linear system of equations, i.e. $A\vec{v} = \vec{b}$, which can be done by the HHL algorithm.

The discretized matrix $A$ is a well-studied Hermitian matrix, of which the eigenvectors are $u_j(k) = \sqrt{2/N} \sin(j\pi k/N)$ and the eigenvalues are $\lambda_j = 4N^2 \sin^2(j\pi/2N)$ [21]. The eigenvalue decomposition of matrix $A$ can be written as $A = S\Lambda S^T$, where $\Lambda$ is a diagonal matrix with elements of the eigenvalues, i.e. $\Lambda = diag(\lambda_1, \lambda_2, ..., \lambda_{N-1})$, and $S$ is an orthogonal matrix composed of the eigenvectors, i.e. $S = [u_1, u_2, ..., u_{N-1}]$. Matrix $S$ is actually the sine transform with the elements of $S_{j,k} = \sqrt{2/N} \sin(\pi jk/N)$. Obviously, $S$ is Hermitian and $S^2 = 1$. Utilizing these properties of matrix $A$, the exponential $A$ can be decomposed as follows,

$$e^{iAt} = S \cdot e^{i\Lambda t} \cdot S. \quad (4)$$

This equation provides us an efficient way to simulate the unitary operator $e^{iAt}$, which is a key step in the HHL algorithm.

The above results for the one-dimensional Poisson equation could be extended immediately to the multidimensional cases. As shown in Ref. [11,21], the discretized matrix $A_d$ for $d$-dimensional Poisson equation can be expressed using the Kronecker products as follows,

$$A_d = \underbrace{A \otimes I \otimes \cdots \otimes I}_{d} + \underbrace{I \otimes A \otimes I \otimes \cdots \otimes I}_{d} + \cdots + \underbrace{I \otimes \cdots \otimes I \otimes A}_{d}. \quad (5)$$

Then the exponential $A_d$ can be written as

$$e^{iA_d t} = \underbrace{e^{iAt} \otimes e^{iAt} \otimes \cdots \otimes e^{iAt}}_{d}. \tag{6}$$

Therefore, the quantum circuit simulating $e^{iA_d t}$ is just the parallel repetitions of the circuit simulating $e^{iAt}$ with the number of $d$. In the following sections, we will mainly focus on the one-dimensional Poisson equation.

## III. DETAILS OF THE QUANTUM CIRCUIT

The present algorithm for solving one-dimensional Poisson equation consists of three main stages as shown in Fig. 1, which are the phase estimation, the controlled rotation and the uncomputation.

The circuit has three main registers, *i.e.* register B, E and A. Register B is used to store the coefficients of the right hand side of Eq. (3). Its number of qubits is $n = \lceil \log(N) \rceil$, where $N$ is the number of discrete points in Eq. (2). Register E is used to store the approximated eigenvalues. Its number of qubits is $m = 2n+2+f$, where the most significant $2n+2$ bits hold the integer part and the remaining $f$ bits the fractional part. The value of $f$ is determined by $f \leq \log \varepsilon_1^{-1}$, where $\varepsilon_1$ is the eigenvalues' error. Register A is used to store approximated angular coefficients for the controlled rotation operation. Its number of qubits is chosen to be equal to that of register B.

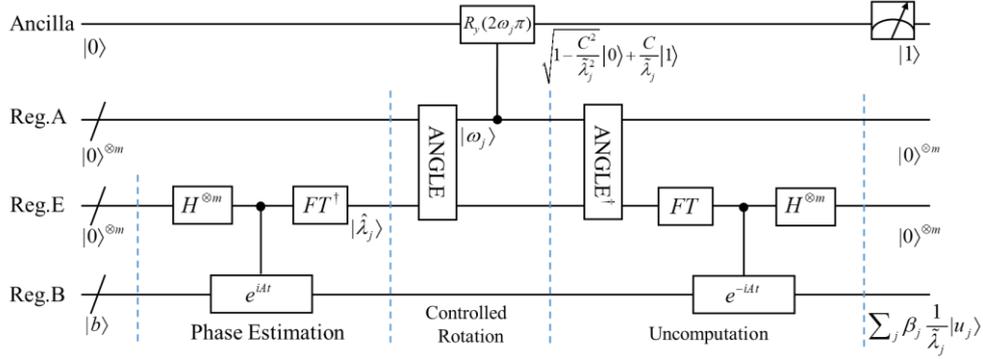

Fig. 1. The overall circuit of the present algorithm for solving one-dimensional Poisson equation. The number of qubits of Reg. B is $n$ and of Reg. E and A is $m$. Note the difference among $\lambda_j$, $\hat{\lambda}_j$ and $\tilde{\lambda}_j$. $\lambda_j$ represents the eigenvalue of matrix $A$. $\hat{\lambda}_j$ denotes the approximation of $\lambda_j$ with precision of $m$ qubits. $\omega_j$ is the angular coefficient evolved from the $\hat{\lambda}_j$. $\tilde{\lambda}_j$ denotes the approximation of $\hat{\lambda}_j$ after the controlled rotation.

Several important caveats to the present algorithm, essentially to the HHL algorithm, should be pointed out here [22]. Firstly, we assume that the input state $|b\rangle$ of Register

B is already available to us. It is prepared as $\sum_i b_i |i\rangle$, where $b_i$ is the values in Eq. (3) and $|i\rangle$ is the computational basis. Several techniques could be used to prepare the state [23-26]. Secondly, the implementation of the circuit could be considered as a process of quantum state preparation. The output of the algorithm is a quantum state which can be written as

$$|v\rangle = A^{-1}|b\rangle = \sum_i \alpha_i |i\rangle, \quad (7)$$

where the solutions of Poisson equation are encoded as the probability amplitudes of the computational basis in the final state.

## A. PHASE ESTIMATION

Phase estimation is used to approximate the eigenvalues of the discretized matrix $A$ and entangle the states encoding the eigenvalues with the corresponding eigenstates [27]. The overall circuit for phase estimation is shown in Fig. 2. Hamiltonian simulation of $e^{iAt}$ is the crucial part in phase estimation.

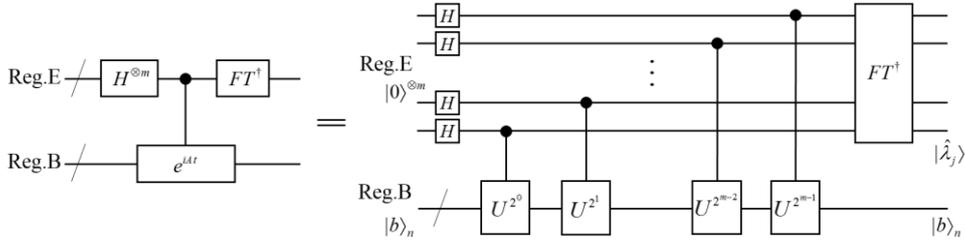

Fig. 2. The circuit of phase estimation. The $U^{2^l}$ represents the unitary operator of $\exp(i2\pi A/2^{m-l})$. $FT^\dagger$ represents the quantum inverse Fourier transform.

Before proceeding to the details of the circuit design, let us first sketch how the quantum states evolve through the circuit. The initial state of register E and B is

$$|0\rangle^{\otimes m}|b\rangle = \sum_{i=1}^{2^n-1} b_i |0\rangle^{\otimes m}|i\rangle = \sum_{j=1}^{2^n-1} \beta_j |0\rangle^{\otimes m}|u_j\rangle, \quad (8)$$

where $|i\rangle$ is the computational basis and $|u_j\rangle$ is the $j$th eigenvector of matrix $A$. Then the Hadamard gates at the start of register E produce a uniformly superposition state. The following sequence of controlled $U^{2^l}$ operation evolves the state as follows,

$$(\sum_{k'=0}^{2^m-1} |k'\rangle\langle k'| ) \otimes U^{k'} \cdot \frac{1}{\sqrt{2^m}} \sum_{k=0}^{2^m-1} |k\rangle \otimes \sum_{j=1}^{2^n-1} \beta_j |u_j\rangle = \sum_{j=1}^{2^n-1} \beta_j \left[ \frac{1}{\sqrt{2^m}} \sum_{k=0}^{2^m-1} \exp(2\pi i \frac{\hat{\lambda}_j}{2^m} k)|k\rangle \right] |u_j\rangle. \quad (9)$$

The state in the square bracket of Eq. (9) is just the output of the quantum Fourier transform acting on the state $|\hat{\lambda}_j\rangle$ [2,28]. So after application of the inverse Fourier

transform, the states evolve into

$$\sum_{j=1}^{2^n-1} \beta_j |\hat{\lambda}_j\rangle |u_j\rangle. \tag{10}$$

The states encoding the approximated eigenvalues, namely $\hat{\lambda}_j$, are prepared and entangled with the eigenstates $|u_j\rangle$.

We now show the details of the circuit for the time evolution of $e^{iAt}$. There are general methods to simulate $e^{iAt}$ [29,30], but here we utilize the specific properties of matrix $A$ to deal with it to reduce the complexity. According to Eq. (4), $e^{iAt}$ can be diagonalized via the sine transform, and the sine transform can be performed based on the $T_N$ and Fourier transform, i.e. $T_N^\dagger F T_{2N} T_N = C_{N+1} \oplus (-iS_{N-1})$ (More details about the sine transform and its quantum implementation can be found in Refs. [11,32].). Fig. 3 and Fig. 4 show the circuits of the controlled $U^{2^0}$ operation and the $T_N$ transform, respectively.

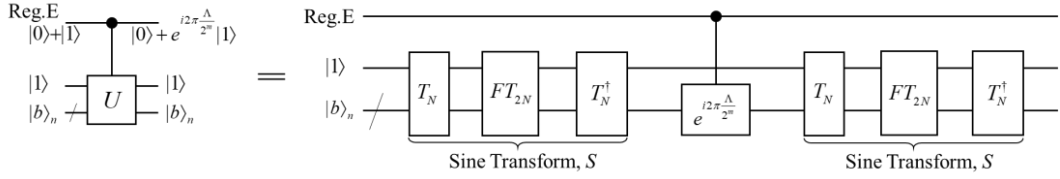

Fig. 3. The circuit of the controlled $U$ operation. One ancillary qubit prepared in state $|1\rangle$ is used to obtain sine transform from the transformation $T_N^\dagger F T_{2N} T_N$.

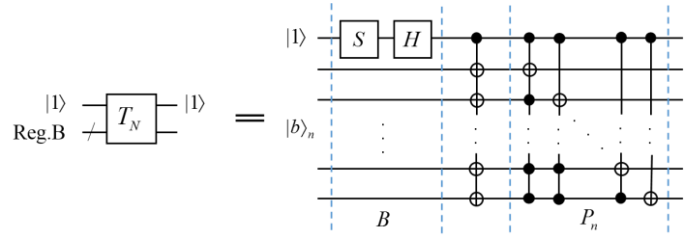

Fig. 4. The circuit of the $T_N$ transform in $T_N^\dagger F T_{2N} T_N$. The present circuit is a simplified one of that in Ref. [32]. The controlled Hadamard and phase gates are ignored because $|b_n\rangle$ has no $|0\rangle^{\otimes n}$ state. The multiple controlled NOT gates could be decomposed into TOFFOLI gates using the method in Ref. [33].

For the operator $\exp(i2\pi \Lambda/2^m)$ in Fig. 3, its eigenvectors are the computational basis, and its eigenvalues are the exponential of matrix $A$'s eigenvalue, namely $\lambda_j$. The action of the operator $\exp(i2\pi \Lambda/2^m)$ on the state $|b\rangle$ of Eq. (8) could be expressed as,

$$\exp(i\Lambda \frac{2\pi}{2^m}) \cdot \sum_{j=1}^{2^n-1} b_j |j\rangle = \sum_{j=1}^{2^n-1} b_j \exp(i\hat{\lambda}_j \frac{2\pi}{2^m}) |j\rangle. \quad (11)$$

This transformation can be implemented using the trick of phase kickback [1]. The general idea of phase kickback is that a computational basis adds a constant integer $\lambda$ mod$2^m$ will kickback a phase change of $\exp(i2\pi\lambda/2^m)$. Fig. 5 shows an example circuit for the controlled $\exp(i2\pi\Lambda/2^m)$ operator with $m$=3. Phase kickback is accomplished using the (inverse) Fourier transformation and the controlled modular-addition operation. A ripple-carry adder (see Ref. [34]) is used to achieve the modular addition operation by omitting the highest carry bit. The ripple-carry adder consists of two basic units, which are SUM and CARRY units. The adder could become a controlled one by changing the CNOT gates of SUM unit into TOFFOLI gates and leaving CARRY unit unchanged.

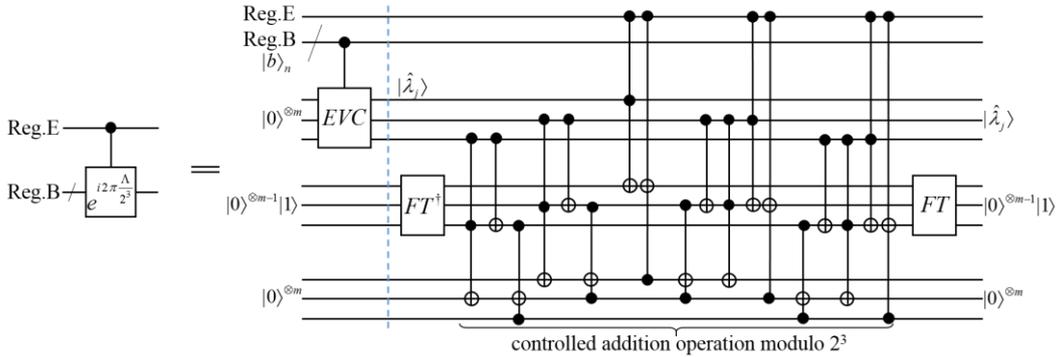

Fig. 5. The circuit of the controlled $\exp(i2\pi A/2^m)$ transformation with $m$=3. The $EVC$ is short for EigenValue Calculation, and this module is used to approximate the eigenvalues. The total number of qubits and operations is $nm+4m$ and $m^3/3+11m^2/2$, respectively.

The module of $EVC$ in Fig. 5 is used to approximate the eigenvalues of matrix $A$ and obtain the state $|\hat{\lambda}_j\rangle$. It is already known that the eigenvalues are $\lambda_j = 4N^2 \sin^2(j\pi/2N)$, i.e. $\lambda_j = 2N^2(1-\cos j\pi/N)$, so the $|\hat{\lambda}_j\rangle$ state could be prepared by computing the cosine function.

Cosine function can be calculated based on the Tayler expansion and repeated squaring method [11], but it need to perform a large number of reversible multiplication and square operations, which require a large number of auxiliary qubits. We try to evaluate the cosine function in a different way which could obtain the binary expansion of the solution in a simple recursive way. The classical counterpart of this method was first proposed by Borwein and Girgensohn, who was inspired by the Plouffe's arctangent identity [18]. For some functions satisfying the requirements of the so-called type (S) system, this method could output the solution value in binary form digit by digit. Each digit of the output binary string is exact, so the error of solutions determined by the number of bits. We refer to this method as Plouffe's binary expansion method.

Inverse trigonometric functions belong to the type (S) system and we will utilize this method to compute arc cotangent function in the controlled rotation as shown in next section. However, cosine function belongs not to the type (S) system. So we revise the Plouffe's binary expansion method for arc cosine function and design the corresponding quantum algorithm to compute the binary solution of cosine function.

The process of computing cosine function is as follows. For $\cos(j\pi/N)$, firstly the binary expansion of the coefficient $j/N = (0.v_{n-1} \cdots v_2 v_1 v_0)_2, v_i \in \{0,1\}$ is prepared in register B before the *EVC* module as shown in Fig. 5. Then the absolute value of the final solution of cosine function, *i.e.* $a_n = |\cos(j\pi/N)|$, can be approximated step-by-step using the following recursive formula,

$$a_0 = 1, \quad a_{i+1} = \begin{cases} \sqrt{(1+a_i)/2}, & \text{if } v_i = 0 \\ -\sqrt{(1-a_i)/2}, & \text{if } v_i = 1 \end{cases}. \tag{12}$$

In other words, the cosine value is approximated step by step based on the value of each qubit in register B.

Fig. 6 shows the circuit of *EVC* module, which is actually used to perform the calculation of $2N^2(1-\cos(j\pi/N))$. The COS modules with the number of *n* are used to implement the recursive operations of Eq. (12) to approximate $\cos(j\pi/N)$. The Adder module is for modular addition operation, which is the same as that in Fig. 5. Note the position of a bar on the right- or left-hand side of Adder network in Fig. 6. The bar on the right-hand side denotes the module is a normal ripple-carry adder, while left-hand side denotes the Adder's elementary gates are in a reversed sequence and it is actually a Subtracter. Calculation are performed in fixed precision arithmetic, so multiplications of 2 and $N^2$ can be performed easily by just keeping in track the position of the decimal point.

The circuit of each COS module is almost the same except with different control qubits. Fig. 7 shows the circuit of the *i*th COS module. The dominant cost of the COS module results from the square root operation.

Bhaskar et al. developed the quantum algorithm for square root operation based on the Newton iteration method [35]. As that for cosine function, this method requires a number of iteration steps and in each iteration step there has multiple multiplication and division operations. We develop a new kind of quantum algorithm for the square root operation based on the non-restoring method. In classical algorithms, the non-restoring method belongs to a kind of digit recurrence method [20]. Like the Plouffe's binary expansion method, this method output the solution in binary form digit-by-digit, of which each bit is the exact one. There are only addition and subtraction operations in each recurrence. In appendix B.1, we describe the computing procedure of this method in detail and give a simple example illuminating the procedure further.

Fig. 8 shows the circuit design for the square root operation based on the non-restoring method. The circuit consists of $2m$ iteration steps, where $m$ is the number of qubits of register E. As can be seen from the figure, each iteration step includes only one subtraction and one controlled addition operation. So the complexity of the present algorithm is $O(m)$ in qubits and $O(m^2)$ in operations, which are the same as that of multiplication operation.

To sum up, the cost of the $EVC$ module, in fact for computing the cosine function, is $O(m^2)$ in qubits and $O(m^3)$ in operations. The complexity of the present method is the same with that in Ref. [11]. However, the present circuit design is much more modular, and the error propagation is easier to control.

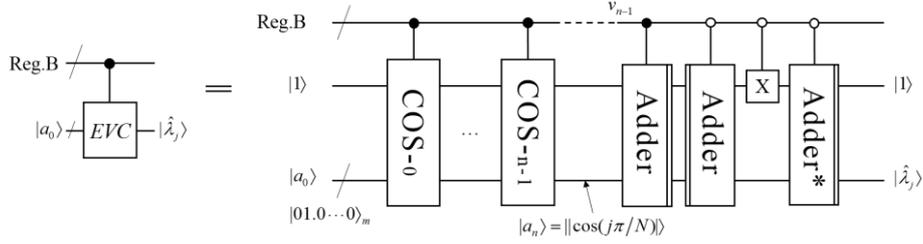

Fig. 6. The overall circuit for eigenvalues computation. Adder denotes a ripple-carry adder omitting the highest carry bit. The ancillary qubit $|1\rangle$ is for addition operation, where for Adder module the augend is designed to be $|1\rangle|0\rangle^{\otimes m}$ and for Adder* be $|0\rangle^{\otimes m}|1\rangle$. The total number of qubits and operations is $m(n+4)$ and $n(33m^2/2+32m)$, respectively.

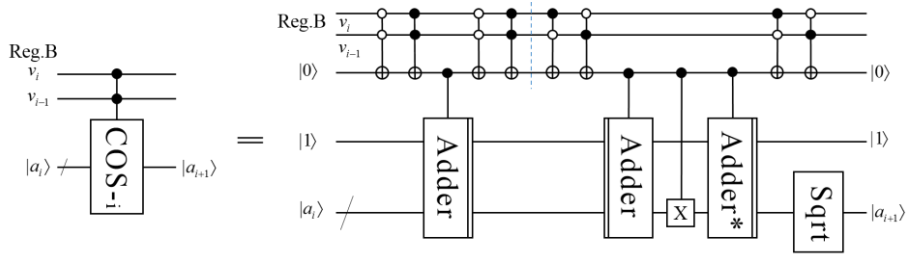

Fig. 7. The circuit for $i$th COS module to accomplish the operation of one iteration in Eq. (12). The total number of qubits and operations is $5m$ and $33m^2/2+32m$, respectively.

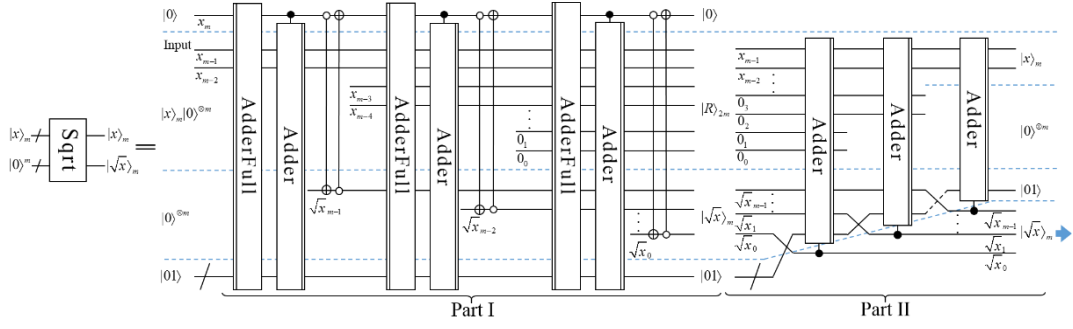

Fig. 8. The circuit for the square root operation. It consists of two parts, where part I is used to calculate the square root and part II is used to uncompute the remainder register $|R\rangle_{2m}$. AdderFull is an unabridged ripple-carry adder [34]. The total number of qubits and operations is $5m$ and $33m^2/2+22m$, respectively.

## B. CONTROLLED ROTATION

After phase estimation, the state of $\sum_{j=1}^{2^n-1} \beta_j |\hat{\lambda}_j\rangle |u_j\rangle$ is obtained in register B and E. Then we perform the linear map taking the state of $|\hat{\lambda}_j\rangle$ to $1/\tilde{\lambda}_j |\hat{\lambda}_j\rangle$, which can be accomplished by the controlled rotation. It consists of two parts: calculating the rotation angular coefficients and performing the controlled $R_y$ operation.

The probability amplitude of $1/\tilde{\lambda}_j$ would be produced by implementing the controlled $R_y$ rotation, namely $R_y(2\theta_j)|0\rangle = \cos\theta_j |0\rangle + \sin\theta_j |1\rangle$. The relationship between rotation angle $\theta_j$ and $\hat{\lambda}_j$ can be expressed as

$$\sin\theta_j = 1/\hat{\lambda}_j. \tag{13}$$

This equation can be further written as,

$$\cot\theta_j = \sqrt{\hat{\lambda}_j^2 - 1}, \theta_j \in (0, \pi/2). \tag{14}$$

Now we omit the subtrahend 1 under the square root of Eq. (14) and take $\theta_j = \omega_j \pi$, so Eq. (14) turns to

$$\omega_j = \frac{\operatorname{arccot}\hat{\lambda}_j}{\pi}, \omega_j \in (0, 1/2). \tag{15}$$

It should be emphasized that the error of $1/\tilde{\lambda}_j$ caused by omitting the subtrahend 1 is less than $2^{-10}$, which will be discussed in detail in appendix C.

As mentioned in the last section, arc cotangent function is a member of the type (S)

system, so Eq. (15) can be solved by the Plouffe's binary expansion method [18]. Specifically, the binary solution of the angular coefficient $\omega_j = (0.w_0 w_1 \cdots w_{m-1})_2$ can be calculated digit by digit using the following recursive formula,

$$w_i = \begin{cases} 0, & \text{if } a_i > 0 \cup \{-\infty\} \\ 1, & \text{if others} \end{cases}; \quad a_0 = \hat{\lambda}_j, \quad a_{i+1} = \begin{cases} 1/2(a_i - 1/a_i), & \text{if } a_i \neq 0 \\ -\infty, & \text{if } a_i = 0 \end{cases}. \quad (16)$$

The overall circuit for the controlled rotation is shown in Fig. 9. The Arccot module in the circuit is designed to implement the operations of Eq. (16). One Arccot module represents one recursive step. Fig. 10 shows the circuit of the $i$th Arccot module. After the $m$ Arccot modules, the $|\omega_j\rangle$ state is prepared in register A. Then the controlled $R_y$ operations take $|0\rangle$ to $\sqrt{1-\left(C/\tilde{\lambda}_j\right)^2}|0\rangle + C/\tilde{\lambda}_j|1\rangle$, where $C$ is a normalizing constant.

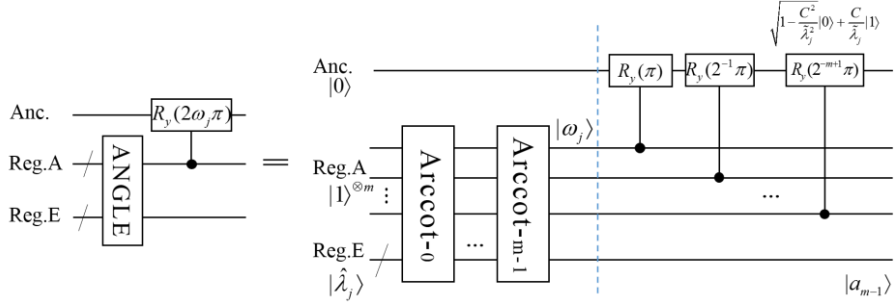

Fig. 9. The circuit of the controlled rotation. The total number of qubits and operations is $m^2+4m$ and $m(34m^2-50m)$, respectively.

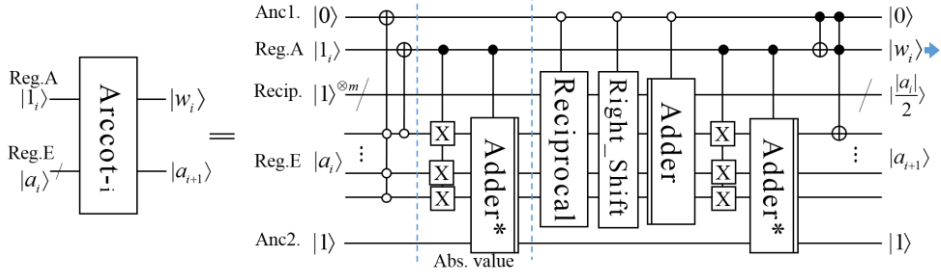

Fig. 10. The circuit for the $i$th Arccot module to compute $\omega_j$. The Adder and Adder* modules are same as those in Fig. 6. The total number of qubits and operations is $4m$ and $34m^2-50m$, respectively.

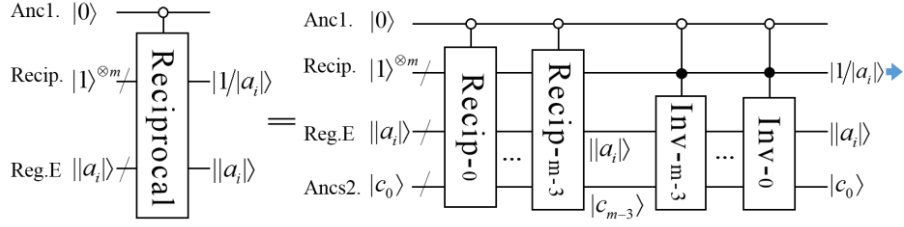

Fig. 11. The circuit for solving the reciprocal function. The initial state of Ancs2. is $|c_0\rangle = |0\rangle^{\otimes m}|1\rangle$. The total number of qubits and operations is $4m$ and $34m^2-68m$, respectively.

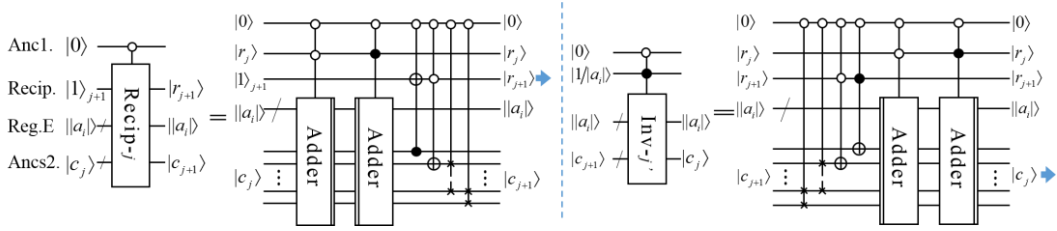

Fig. 12. The circuits of Recip-$j$ and Inv-$j'$ modules. The Adder module is the same as that in Fig. 6. The total number of qubits and operations is $3m$ and $34m$, respectively.

As shown in Fig. 10, the Arccot module contains reciprocal operation. Bhaskar et al. also discussed the quantum algorithm for reciprocal operation based on Newton iteration method [35]. Like that for the square root operation, we develop a quantum digit-recurrence algorithm for the reciprocal operation based on the non-restoring method [19]. The computing procedure of this method, as well as an illustration example is described in appendix B.2.

Fig. 11 shows the quantum circuit for the reciprocal operation based on the non-restoring method. It consists of $m$-2 Recip modules used to calculate the reciprocal value and $m$-2 Inv modules to uncompute the ancillary register. The circuits of the Recip and Inv modules are shown in Fig. 12. There are only one subtraction and one addition operation in the Recip and Inv modules. The cost of the present algorithm for computing reciprocal is $O(m)$ in qubits and $O(m^2)$ in operations, which are the same as that of multiplication operation. The binary output has the precision of up to $m$ qubits.

To sum up, the complexity of the present algorithm for the controlled rotation operation, in fact for computing the arc cotangent function, is $O(m^2)$ in qubits and $O(m^3)$ in operations, which is the same with the *EVC* module. In Ref. [11], the complexity is $O(m^3)$ qubits and $O(m^4)$ operations. So our method reduces the complexity by one order. Furthermore, the present circuit design is much more modular, and the error propagation is easier to control.

After controlled rotation, the uncomputation is implemented to evolve the state of registers B, E and A back to the initial state. Finally, we perform the measurement operation. If the measurement result of the ancillary qubit is $|1\rangle$, then we know the final state is $\sum_j \beta_j \tilde{\lambda}_j^{-1} |u_j\rangle$ which is proportional to $A^{-1}|b\rangle$, namely the solution state for the Poisson equation.

## D. ALGORITHM COMPLEXITY AND ERROR ANALYSIS

The present algorithm for solving one-dimensional Poisson equation takes the HHL algorithm as the template as shown in Fig. 1. HHL uses roughly $O(\kappa^2 \log(N)/\varepsilon)$ steps to output the solution state with a success probability arbitrarily close to one, where $\varepsilon$ is the error of the solution state and $\kappa$ and $N$ is the discretized matrix's condition number and dimension, respectively [5]. Since we utilize the specific properties of the discretized matrix $A$ to simulate the Hamiltonian $e^{iAt}$, the complexity of the present algorithm should be lower than that of HHL.

Table 1 shows the number of qubits and elementary gates used in each part of our circuit as shown in Fig. 1. The elementary gates refer to the CNOT, TOFFOLI and single qubit gates. As can be seen from the table, the total cost of the circuit is $O(m^2)$ qubits and $O(m^3)$ operations, where $m$ is the number of qubits of register E and A. Note this is the cost for one computation. Using the trick of amplitude amplification [36], a number of repetitions proportional to $\kappa$ could lead to a success probability arbitrarily close to one [5]. So the complexity of the present algorithm for solving one-dimensional Poisson equation is $O(m^2)$ qubits and $O(\kappa m^3)$ operations.

Tab. 1. The number of qubit and elementary gates of each modules of our algorithm

| Modules | | Qubits [a] | Elementary Gates [a] |
|---|---|---|---|
| Phase Estimation | $S$ [b] | $n+2$ | $97n^2-745n$ |
| | EVC | $m(n+4)$ | $33nm^2+64nm$ |
| | Phase Kickback | $3m$ | $m^3/3+11m^2/2$ |
| Controlled Rotation | Angle Computing | $m^2+3m$ | $34m^3-50m^2$ |
| | Controlled $R_y$ | $m+1$ | $4m$ |
| Uncomputation | | $m^2+nm$ | $34m^3+33nm^2$ |
| Total | | $m^2+nm$ | $68m^3+66nm^2$ |

[a] Only the first two significant terms are written for simplicity.

[b] The Sine transform accomplished through the $T_N^\dagger F T_{2N} T_N$ transformation.

The condition number $\kappa$ of the discretized matrix $A$ is *polynomial*($N$) (see appendix A for details), so the present algorithm could not offer exponential speedup over classical algorithms for solving one-dimensional Poisson equation. However, for $d$-dimensional Poisson equation the present algorithm achieves exponential speedup in terms of $d$. In fact, the main difference of the circuits between $d$- and one- dimensional case is the Hamiltonian simulation in the phase estimation. According to Eq. (6), the circuit simulating $e^{iA_d t}$ can be obtained by the replication and parallel application of the circuit simulating $e^{iAt}$ used in one-dimensional case. So the complexity of the present algorithm for solving $d$-dimensional Poisson equation is $O(dm^2)$ qubits and $O(\kappa dm^3)$ operations, which is linear with $d$. In terms of error $\varepsilon$, the complexity can be further expressed as follows,

$$\begin{cases} O\left(d\log^2(\varepsilon^{-\alpha})\right) & qubits \\ O\left(\kappa d\log^3(\varepsilon^{-\alpha})\right) & operations \end{cases}. \qquad (17)$$

This equation uses the following two facts: first, $m = \Theta(n) = \Theta(\log(N))$; second, the basic error is from the central-difference approximation of the second derivative (see Eq. (2)), and it satisfies $N = \varepsilon^{-\alpha}$, where α>0 is a constant depending on the smoothness of the solution. Any direct or iterative classical algorithm solving the $d$-dimensional Poisson equation has a cost of at least $\varepsilon^{-\alpha d}$, which increases exponentially with the dimension [37]. On the other hand, the complexity of Cao's algorithm is $\max\{d,\log(\varepsilon^{-\alpha})\}\cdot\log^2(\varepsilon^{-\alpha})$ in qubits and $\max\{d,\log(\varepsilon^{-\alpha})\}\cdot\log^3(\varepsilon^{-\alpha})$ in operations. In the cases of low dimensions or high precisions, the present algorithm's complexity is lower than that of Cao's by one order.

The variation trends of the complexities mentioned above are summarized in Fig. 13. In the diagram, we assume that $\varepsilon = 2^{-23}$, namely the single-precision, and use the fact that $\kappa = \varepsilon^{-2\alpha}$ (see Appendix A). One remarkable point is that the speedup of present algorithm against the classical one appears when the dimension of the Poisson equation reaches about three.

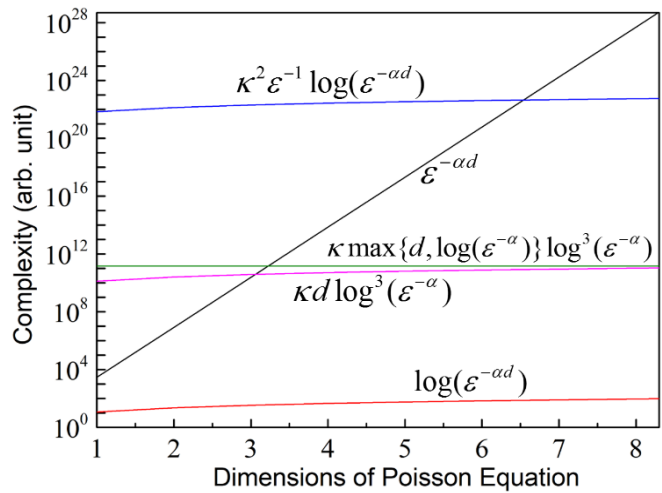

Fig. 13. The sketch of the complexity of different algorithms for solving the $d$-dimensional Poisson equation. From top to bottom, the formulas represent, respectively, the complexity of HHL, classical algorithm, Cao's algorithm, the present algorithm and the exponential speedup of classical algorithm.

Now we analyze the error of the final solution state $\sum_{j=1}^{2^n-1} \beta_j \tilde{\lambda}_j^{-1} |u_j\rangle$, in fact the error of the probability amplitude $\tilde{\lambda}_j^{-1}$. We approximate the reciprocal of eigenvalues through two stages. First, we obtain $\hat{\lambda}_j$ by approximating $\lambda_j$ in the phase estimation, and then we obtain $\tilde{\lambda}_j^{-1}$ from $\hat{\lambda}_j$ in the controlled rotation. So the error can be calculated by the following expression,

$$\left| \frac{1}{\tilde{\lambda}_j} - \frac{1}{\lambda_j} \right| = \left| \left( \frac{1}{\hat{\lambda}_j} - \frac{1}{\lambda_j} \right) + \left( \frac{1}{\tilde{\lambda}_j} - \frac{1}{\hat{\lambda}_j} \right) \right|. \tag{18}$$

For the first item, we have

$$\frac{1}{\hat{\lambda}_j} - \frac{1}{\lambda_j} = \frac{\lambda_j - \hat{\lambda}_j}{\hat{\lambda}_j \lambda_j} < \frac{2^{-f}}{(\hat{\lambda}_j)^2} \in [0, 2^{-f-6}], \tag{19}$$

where $f$ is the number of qubits holding the fractional part in register E. The above equation use the fact that $\hat{\lambda}_j$ is no less than 8. In fact, the smallest eigenvalue is $\lambda_1(N) = 4N^2 \sin^2(\pi/2N)$ and it is a monotonic increasing function with $N$. So we have the inequality of $\lambda_1(N) \geq \lambda_1(2) = 8, N \geq 2$. Using the same fact, the second item would be expressed as

$$\frac{1}{\tilde{\lambda}_j} - \frac{1}{\hat{\lambda}_j} = \frac{1}{\sqrt{1+\hat{\lambda}_j^2}} - \frac{1}{\hat{\lambda}_j} \in (-2^{-10}, 0), \tag{20}$$

which results from the omitting of the subtrahend 1 in Eq. (14).

The total error is $|2^{-f-6} - 2^{-10}|$, where the second item is the main limit. However, This error can be reduced further by amplifying the eigenvalues, *i.e.* shift the binary string of $\hat{\lambda}_j$ left. More details about the deduction of Eq. (20) and the method of reducing the error can be seen in appendix C.

## IV. CIRCUITS DEMONSTRATION ON A QUANTUM VIRTUAL SYSTEM

We demonstrate our algorithm on a quantum virtual computing system installed on the Sunway TaihuLight supercomputer of the National Supercomputing Center in Wuxi. This virtual system was developed by the Origin Quantum Computing Technology co., LTD in Hefei of China. It has three operating modes, which are Full Amplitude mode, Partial Amplitude mode and Single Amplitude mode, respectively. The Full Amplitude mode could output the probabilities of all the final states for a circuit with at most 43 qubits. The Partial Amplitude mode could output the amplitudes of partial final states for a 82 qubits circuit, while the Single Amplitude mode the amplitude of one final state for a 200 qubits circuit with at most 20 depths [38].

Because of the limitation of the quantum virtual machine and the computing resource, we demonstrate our algorithm in two steps. Firstly, we propose a simplified version of our circuit for the one-dimensional Poisson equation with $N = 4$ discretized points to demonstrate the general effectiveness of our algorithm. The circuit is shown in Fig. 14. Since the *EVC* (Fig. 5) and *ANGLE* modules (Fig. 9) contribute the most complicated parts of our algorithm, in the simplified circuit the two modules are replaced by the modules which would prepare directly the eigenvalues and the rotation angular coefficients into the register E and A, respectively. Note the operation cost of such modules grows exponentially with the number of qubits, so they could not be used in real circuit.

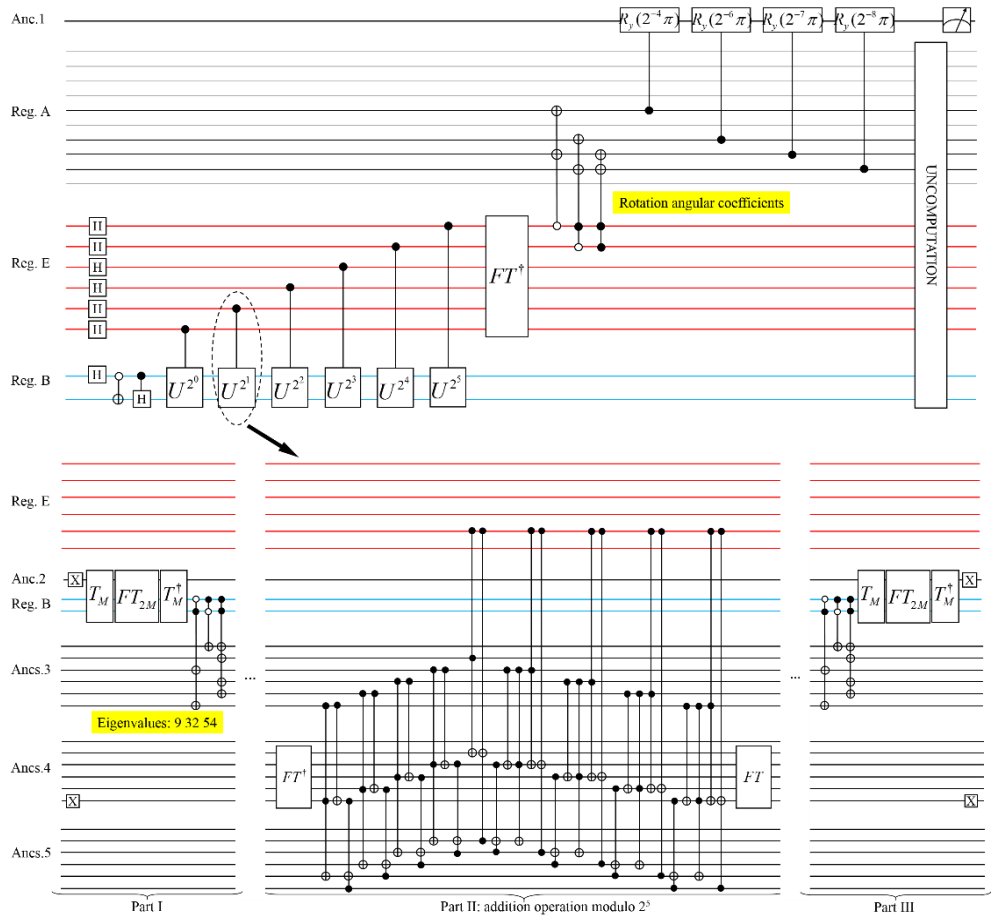

Fig. 13. The simplified circuit of our algorithm for solving Poisson equation. The upper subgraph shows the overall circuit, while the lower one shows the circuit of controlled $U^{2^1}$ transformation. The gray lines in Register A denotes that such qubits could be omitted. The first three operations in Register B are used to prepare the initial state of $|b\rangle = 1/\sqrt{2}|01\rangle + 1/2|10\rangle + 1/2|11\rangle$. The circuit for $T_N$ transformation is shown in Fig. 4.

Secondly, the circuits for the cosine, arc cotangent, square root and reciprocal functions (as shown in Figs. 7, 10, 8, 11, respectively) are demonstrated individually. The configuration parameters of implementing the circuits on the quantum virtual computing system are listed in Tab. 2. All the codes of the circuits are provided as the attachment files of the present paper, which can be found in Ref. [39]. The codes are written using the quantum assembly language of QRunes developed by the Origin Quantum Computing Company [40].

Tab. 2. The configuration parameters of executing our circuits on the quantum virtual system

|  | Poisson[a] | Cosine | Arc cot | Square Root | Reciprocal |
|---|---|---|---|---|---|
| Qubits | 38 | 31 | 24 | 26 | 21 |
| Gates | 900 | 800 | 1100 | 350 | 800 |
| Measure Times | 1 | 1 | 1 | 1 | 1 |
| Least Nodes[b] | 1024 | 8 | 1 | 1 | 1 |
| Operation Mode | Full Amp. | Full Amp. | Full Amp. | Full Amp. | Full Amp. |
| Operation Hours | 0.35 | 0.55 | 0.15 | 0.10 | 0.10 |

[a] The simplified circuit of the present algorithm as shown in Fig. 13.
[b] One node of the Sunway TaihuLight supercomputer contains 260 computing cores.

The running results of our circuits, including the Poisson, cosine, arc cotangent, square root and reciprocal functions, are listed in Tab. 3. For the Poisson circuit, the input is an entanglement state encoding the coefficients of the right-hand side of the Poisson equation as the probability amplitudes, and the solutions are encoded into the probability amplitudes of the output states as indicated in Eq. (7). The real solution of the sample Poisson equation is (0.9053, 1.1036 0.8018), which after normalization turns to (0.553 0.674 0.490). So the error of the output solutions is less than 0.5%. The success probability of obtaining the desired state in one computation is 1.12%, which is proportional to $\kappa^{-2}$. If not use the amplitude amplification trick, $O(\kappa^2)$ repetitions of computations are needed to achieve a success probability arbitrarily close to 1.

The Cosine circuit evaluate the function of $\cos(j\pi/N)$ with $N = 4$, and the $j$ are encoded as the input states. The binary string of the output states represents the complement of the solution, namely when the most significant bit is 1, you need invert each bit and plus one to obtain the final solution. The Arc-cot, Square Root and reciprocal circuits evaluate the function of $\text{arccot}(x)/\pi$, $\sqrt{x}$ and $1/x$, respectively. As can be seen from the results, each bit of the binary string of the solution is exact and the error is determined totally by the number of the qubits. All the running results verify the effectiveness of our algorithms.

Tab. 3. The results of our circuits implemented on the quantum virtual system

| | Input States | Output States |
|---|---|---|
| Poisson [a] | $\frac{1}{\sqrt{2}}|01\rangle + \frac{1}{2}|10\rangle + \frac{1}{2}|11\rangle$ | $0.551|01\rangle + 0.675|10\rangle + 0.491|11\rangle$ |
| Cosine | $|00\rangle, |01\rangle, |10\rangle, |11\rangle$ | $|01.000\rangle, |00.101\rangle, |00.000\rangle, |11.011\rangle$ |
| Arc-cot | $|01.00\rangle$ | $|.01\rangle$ |
| Square Root | $|0010\rangle, |0111\rangle, |1001\rangle, |1111\rangle$ | $|01.01\rangle, |10.10\rangle, |11.00\rangle, |11.11\rangle$ |
| Reciprocal | $|0010\rangle, |0011\rangle, |1000\rangle, |1111\rangle$ | $|0.1000\rangle, |0.0101\rangle, |0.0010\rangle, |0.0001\rangle$ |

[a] The Poisson circuit is implemented on Partial Amplitude mode, others on Full Amplitude mode.

## V. CONCLUSIONS

In the present work, we develop a quantum Fast Poisson Solver, including the algorithm and the complete and modular circuit design. The algorithm is generally based on the HHL algorithm and it achieves an exponential speedup in terms of dimension of Poisson equation. We perform the controlled rotation based on the arc cotangent function, so the rotation angles are prepared directly from the eigenvalues, instead of its reciprocals. The Plouffe's binary expansion method are used to evaluate the arc cotangent and cosine function. Based on the cosine function, we perform the eigenvalue approximation for the Hamiltonian simulation in a different way. In addition, Quantum algorithms for solving square root and reciprocal functions are developed based on the non-restoring digit-recurrence method. All these developments make the present algorithm have lower complexity and the circuit design complete and modular. Comparing with Cao's algorithm, the present one reduces the complexity by one order in the case of low dimension of Poisson equation or high precision of solutions. And the speedup of our algorithm against the classical method appears when the dimension of Poisson equation reaches about three.

The effectiveness of the present circuits has been demonstrated on a quantum virtual computing system installed on the Sunway TaihuLight supercomputer. We think this is an important step towards the real applications of the quantum algorithm for solving Poisson equation as a Fast Poisson Solver in the near-term hybrid classical/quantum system. For the next step, we will study how to embed the quantum Fast Poisson Solver into the classical program for practical applications.

## ACKNOWLEDGMENTS

We are very grateful to the National Supercomputing Center in Wuxi for the great computing resource. We would also like to thank the technical team from the Origin Quantum Computing Technology co., LTD in Hefei for the professional services on quantum virtual computation. The present work is financially supported by the National Natural Science Foundation of China (Grant No. 61575180, 61701464, 11475160) and the Pilot National Laboratory for Marine Science and Technology (Qingdao).

# APPENDIX
## A. Condition number of the discretized matrix

We take the maximum eigenvalue as the norm of the discretized matrix $A$. So the norm of matrix $A$ is $\lambda_{max} = 4N^2 \sin^2 \frac{(N-1)\pi}{2N}$ as shown in Sec. II, and that of matrix $A^{-1}$ is $\lambda'_{max} = (4N^2 \sin^2 \frac{\pi}{2N})^{-1}$. The condition number of matrix $A$, or the ratio between $A$'s largest and smallest eigenvalues [5], can be obtained

$$\kappa = \|A\| \cdot \|A^{-1}\| = \lambda_{max} \cdot \lambda'_{max} = \cot^2 \frac{1}{2N} \pi. \tag{A.1}$$

The relationship between $\kappa$ and $N$ is apparently nonlinear. Assuming $x = \pi/2N < 1$, the squared cotangent function satisfies the following equalities,

$$\cot^2 \frac{1}{2N} \pi = \cot^2 x = \left( i \frac{e^{ix} + e^{-ix}}{e^{ix} - e^{-ix}} \right)^2$$
$$= \frac{1}{x^2} - 1 + o(x^3) < \frac{N^2}{2} - 1 + o((\frac{\pi}{2N})^3), \tag{A.2}$$

where Taylor expansion of $e^{ix}$ and $e^{-ix}$ are used. So the condition number of matrix $A$ is $\kappa = O(N^2)$. In addition, we have the relationship between $N$ and the basic error of the solutions caused by the central-difference approximation, namely $N = \varepsilon^{-\alpha}$. The $\alpha$ is a smoothness parameter depending on the smoothness of the solution function. (For example, when the solution function has uniformly bounded partial derivatives up to order four, $\alpha$ is 1/2 [21].) Therefore, the condition number of matrix $A$ can be expressed as $\kappa = O(N^2) = O(\varepsilon^{-2\alpha})$, which is independent of the dimension of matrix $A$. The additive preconditioner [41] would be used to reduce the $\kappa$.

## B. non-restoring method for the square root and reciprocal operation
B.1. Square root operation

The calculation procedure consists of the following six steps.

1st. Ignore the radix point of the binary number and Expand the $m$ bits binary string to be $2m$ bits by adding 0 on the right hand side.

2nd. Divide the $2m$ bits string into $m$ parts in pairs from upper bit to lower.

3rd. Subtract 01 from the most left part. If the subtraction result is non-negative, then the first bit of the solution is 1 and proceed to the next step. Otherwise, first bit of the solution is 0 and undo the subtraction operation.

4th. Expand the subtraction result of the last step by combining it with the next 2-bits part, and combine the solution obtained in the last step with 01. Then subtract the second number from the first number. If the subtraction result is non-negative, the next bit of the solution is 1, and proceed to the next step. Otherwise, the next bit is 0, and undo the subtraction operation.

5th. Repeat Step 4 *m*-1 times and the *m* bits of the solution is obtained.

6th. Performing right shift operation to get result.

We take the square root of $01.00_2$ as the example to illustrate the calculating procedure as shown in Fig. B.1.

$$
\begin{array}{r}
\phantom{\sqrt{0}} \text{I \ II \ III \ IV} \\
\phantom{\sqrt{0}} \hat{1}.\tilde{0}\ \hat{0}\ 0 \\
\sqrt{01.00\ 00\ 00} \\
-01 \phantom{0000000} \\
\hline
\phantom{-}00\ 00 \phantom{0000} \\
-\hat{1}\ 01 \phantom{0000} \\
\hline
\phantom{-}00\ 00\ 00 \phantom{00} \\
-\hat{1}\tilde{0}\ 01 \phantom{00} \\
\hline
\phantom{-}00\ 00\ 00 \\
-\hat{1}\ \tilde{0}\tilde{0}\ 01 \\
\hline
\cdots\ \cdots
\end{array}
$$

← Result, shift down to $01.00_2$.

← I. Non-negative, the result's first bit is 1,

← II. Negative, the result's second bit is 0, undo,

← III. Negative, the result's third bit is 0, undo,

← IV. Negative, the result's fourth bit is 0, undo.

Fig. B.1. The demo case of calculating square root of $01.00_2$ using our revised non-restoring square root method.

B.2. Reciprocal operation

The calculation procedure consists of the following three steps.

1st. Let the highest bit of dividend be the sign bit. Subtract the divisor from the Dividend. If the sign bit of the subtraction result is 0, the first bit of the quotient solution is 1, otherwise the first bit is 0. Then let the subtraction result shift left one bit except the sign bit with zero padding to the end.

2nd. If the sign bit of the subtraction result is 0, subtract the divisor from the subtraction result of the last step. If the sign bit is 0, add the divisor to the subtraction result.

3rd. Repeating Step 1 and 2 *m* times and the *m* bits of the quotient solution is obtained.

The reciprocal of $1000_2$ is taken as the example as describe in Fig. B.2.

$$
\begin{array}{r}
\hat{0}0001 \\
-1000 \\
\hline
\hat{1}1001 \\
\hat{1}0010 \\
+1000 \\
\hline
\hat{1}1010 \\
\hat{1}0100 \\
+1000 \\
\hline
\hat{1}1100 \\
\hat{1}1000 \\
+1000 \\
\hline
\hat{0}0000 \\
\hat{0}0000 \\
-1000 \\
\hline
\cdots\cdots
\end{array}
$$

← The first calculation is always subtraction,

← The sign bit of remainder is 1, the first bit of quotient is 0,

← Left shift, the sign bit unchanged, pad 0. The next calculation is addition,

← The sign bit of remainder is 1, the second bit of quotient is 0,

← The sign bit of remainder is 1, the third bit of quotient is 0,

← The sign bit of remainder is 0, the fourth bit of quotient is 1,

← Left shift, the sign bit unchanged, pad 0. The next calculation is subtraction.

Fig. B.2. The demo case of calculation reciprocal of $1000_2$ using the non-restoring method.

C. Error Reduction

According to Eq. (13)-(14), the probability amplitude of the quantum state after the controlled rotation operation should be,

$$\sin\theta_j = \frac{1}{\sqrt{1+\cot^2\theta_j}} = \frac{1}{\hat{\lambda}_j}, \theta_j \in (0, \pi/2). \tag{C.1}$$

Substituting Eq. (15) into Eq. (C.1), then it turns to

$$\sin\theta'_j = \frac{1}{\sqrt{1+\hat{\lambda}_j^2}}, \theta'_j \in (0, \pi/2). \tag{C.2}$$

Based on Eq. (15), we know that $1/\tilde{\lambda}_j = 1/\sqrt{1+\hat{\lambda}_j^2}$. So the error is

$$\varepsilon_2(0) = |\sin\theta'_j - \sin\theta_j| = \left|\frac{1}{\sqrt{1+\hat{\lambda}_j^2}} - \frac{1}{\hat{\lambda}_j}\right| < 2^{-10}. \tag{C.3}$$

We can reduce error by amplifying the approximated eigenvalues, *i.e.* shift the binary numeral of $\hat{\lambda}_j$ left, say, $i$ bits. Now the error turns to

$$\begin{aligned}
\varepsilon_2(i) &= \left|\frac{1}{\tilde{\lambda}_j} - \frac{1}{\hat{\lambda}_j}\right| = \left|\frac{1}{\sqrt{2^{-2i}+\hat{\lambda}_j^2}} - \frac{1}{\hat{\lambda}_j}\right| \le \left|\frac{1}{\sqrt{2^{-2i}+8^2}} - \frac{1}{8}\right| \\
&= 2^{-3}\left(2^0 - \frac{1}{\sqrt{2^0+2^{-2i-6}}}\right) \underset{NRM}{<} 2^{-3}\left(2^0 - \frac{1}{2^0+2^{-2i-7}}\right) \\
&\underset{NRM}{=} 2^{-3}\left(2^0 - \sum_{j=1}^{i} 2^{-2j-7}\right) = 2^{-2i-10},
\end{aligned} \tag{C.4}$$

where NRM means the operation is calculated by non-restoring method. So the upper limit of error $\varepsilon_2(i)$ reduces exponentially with $i$. Shifting one bit left make the error reduce about $2^2$ times. At the same time, the state after controlled rotation operation turns to $\sqrt{1-\left(C'/2^i\tilde{\lambda}_j\right)^2}|0\rangle + C'/2^i\tilde{\lambda}_j|1\rangle$, where $C'$ represents the normalizing constant. Factor $1/2^i$ will be contained in the normalizing constant.

## REFERENCES

[1] R. Cleve, A. Ekert, C. Macchiavello, and M. Mosca. Proc. R. Soc. Lond. A **454**, 339-354 (1998).
[2] M. A. Nielsen and I. L. Chuang. *Quantum computation and quantum information*. (Cambridge University Press, Cambridge, 2010).
[3] P. W. Shor. In *Proceedings 35th annual symposium on foundations of computer science* (IEEE, New York, 1994), pp. 124-134.
[4] L. Grover. Phys. Rev. Lett. **79**, 325-328 (1997).
[5] A. W. Harrow, A. Hassidim, and S. Lloyd. Phys. Rev. Lett. **103**, 150502 (2009).
[6] J. Biamonte, P. Wittek, N. Pancotti, P. Rebentrost, N. Wiebe and S. Lloyd. Nature **549**, 195-202 (2017).
[7] S. K. Leyton and T. J. Osborne. arXiv: 0812.4423.
[8] W. Berry. J. Phys. A: Math. Theor. **47**, 105301 (2014).
[9] W. Berry, A. M. Childs, A. Ostrander and G. Wang. Commun. Math. Phys. **356**, 1057-1081

(2017).

[10] J. M. Arrazola, T. Kalajdzievski, C. Weedbrook and S. Lloyd. Phys. Rev. A **100**, 032306 (2019).

[11] Y. Cao, A. Papageorgiou, I. Petras, J. Traub, and S. Kais. New J. Phys. **15**, 013021, (2013).

[12] A. Kosior and H. Kudela. Computers and Fluids **80**, 423-428 (2013).

[13] R. Steijl and G. N. Barakos. Computers and Fluids **173**, 22-28 (2018).

[14] R. W. Hockney. Journal of the Association for Computing Machinery **12**, 95-113 (1965).

[15] B. L. Buzbee, G. H. Golub, and C. W. Nielson. SIAM Journal on Numerical analysis **7**, 627-656 (1970).

[16] P. N. Swarztrauber. SIAM Review **19**, 490-501 (1977).

[17] Y. Cao, A. Daskin, S. Frankel, and S. Kais. Molecular Physics **110**, 1675-1680 (2012).

[18] J. M. Borwein and R. Girgensohn. Can. J. Math. **47**, 262-273 (1995).

[19] C. V. Freiman. *Proceedings of the IRE* **49**, 91-103 (1961).

[20] T. Sutikno. International journal of computer theory and engineering **3**, 46 (2011).

[21] J. W. Demmel. *Applied numerical linear algebra*. (SIAM, Philadelphia, 1997), chap. 6.

[22] S. Aaronson. Nature Physics **11**, 291-293 (2015).

[23] Biham, O. Biham, D. Biron, M. Grassl, and D. A. Lidar. Phys. Rev. A **60**, 2742 (1999).

[24] L. K. Grover. Phys. Rev. Lett. **85**, 1334 (2000).

[25] L. Grover and T. Rudolph. arXiv:quant-ph/0208112

[26] A. N. Soklakov and R. Schack. Phys. Rev. A **73**, 012307 (2006).

[27] A. Luis and J. Peřina. Phys. Rev. A **54**, 4564 (1996).

[28] Y. S. Weinstein, M. A. Pravia, E. M. Fortunato, S. Lloyd, and D. G. Cory. Phys. Rev. Lett. **86**, 1889 (2001).

[29] S. Lloyd. Science **273**, 1073-1078 (1996).

[30] D. W. Berry, G. Ahokas, R. Cleve, and B. C. Sanders. Math. Phys. **270**, 359-371 (2007).

[31] M. V. Wickerhauser. *Adapted wavelet analysis: from theory to software*. (A. K. Peters/CRC Press, Wellesley, 1994) chap. 3.

[32] A. Klappenecker and M. Rotteler. arXiv: quant-ph/0111038.

[33] A. Barenco, C. H. Bennett, R. Cleve, D. P. DiVincenzo, N. Margolus, P. Shor, T. Sleator, J. A. Smolin, and H. Weinfurter. Phys. Rev. A **52**, 3457 (1995).

[34] V. Vedral, A. Barenco, and A. Ekert. Phys. Rev. A **54**, 147 (1996).

[35] M. K. Bhaskar, S. Hadfield, A. Papageorgiou, and I. Petras. Quantum Inf. Comput. **16**, 197-236 (2016).

[36] G. Brassard, P. Høyer, M. Mosca and A. Tapp. Contemporary Mathematics Volume **305** (2002). arXiv: quant-ph/0005055v1.

[37] K. Ritter and G. W. Wasilkowski. Lectures in Applied Mathematics Volume **32** (1996).

[38] Zhao-Yun Chen, Qi Zhou, Cheng Xue, Xia Yang, Guang-Can Guo, and Guo-Ping Guo. Science Bulletin **63**, 964-971 (2018).

[39] The web address for accessing our QRunes codes.

[40] User manual of QRunes, "https://github.com/OriginQ/QPanda/tree/master/QRunes".

[41] V. Y. Pan, D. Ivolgin, B. Murphy, R. E. Rosholt, Y. Tang, and X. Yan. Linear Algebra and Its Applications **432**, 1070-1089 (2010).